\documentclass[referee]{raa}
\usepackage{graphicx,times}
\usepackage{natbib}
\usepackage{amssymb,amsmath}
\usepackage{placeins}
\bibpunct{(}{)}{;}{a}{}{,}

\usepackage[a4paper=true,dvipdfm=true,pagebackref=true]{hyperref}
\hypersetup{pdftitle = The title of my PDF, pdfauthor = My name, pdfsubject= The subject, pdfkeywords = keyword1 keyword2 keyword3}
\hypersetup{colorlinks = true, linkcolor = green, anchorcolor = red, citecolor = blue, filecolor = red, pagecolor = red, urlcolor = red}

\begin{document}

    \title{Moon night sky brightness simulation for Xinglong station
}

 \volnopage{ {\bf 2012} Vol.\ {\bf X} No. {\bf XX}, 000--000}
   \setcounter{page}{1}

   \author{
   Song Yao\inst{1,2} 
   \and 
   Hao-Tong Zhang\inst{2}\footnote{Corresponding author}
   \and
    Hai-Long Yuan\inst{2}
    \and Yong-Heng Zhao\inst{2}
    \and Yi-Qiao  Dong\inst{2}
    \and Zhong-Rui  Bai\inst{2}
    \and Li-Cai Deng\inst{2}
    \and Ya-Juan  Lei\inst{2}
   }

\institute{Graduate University of Chinese Academy of Sciences, Beijing 100049, China;\and
    Key Laboratory of Optical Astronomy, National Astronomical Observatories, Chinese Academy of Sciences, 20A Datun Road, Chaoyang District, Beijing, China, 100012; {\it  email:  zht@lamost.org}\\
    {\small Received ; accepted  }
}

\abstract{With a sky brightness monitor  in Xinglong station of National Astronomical Observatories of China (NAOC), we collected data from 22 dark clear nights and 90  lunar nights. We first measured the sky brightness variation  with time
 in  dark nights, found a clear correlation between the sky brightness and human activity.  Then with a modified sky brightness model of moon night and data from moon night, we derived the typical value for several important parameters in the model. With these results, we calculated the sky brightness distribution under a given moon condition for Xinglong station. Furthermore, we simulated the moon night sky brightness distribution in a 5 degree field of view telescope (such as LAMOST).  These simulations will be helpful to determine the magnitude limit,  exposure time  as well as the survey design for LAMOST at lunar night.
}

   \authorrunning{S. Yao, et. al. }            
   \titlerunning{Moon night sky brightness simulation for Xinglong station}  
   \maketitle

%

\section{Introduction}
\label{sect:intro}

The night sky brightness of the observatory, which is not completely dark, is the major factor that constrains the limiting magnitude  of a telescope and the exposure time given  the demanded signal to noise ratio of the targets. For a fiber spectral instrument, sky subtraction  is the crucial step in fiber spectra data reduction,
 usually this step relies on how the fiber sample the background which is generally thought to be homogenous within 1 degree in  dark nights. Once there is a gradient in sky background (such as
 in moon night), much more effort will be needed in both designing the sky sampling fiber  and data reduction. Thus a good estimation of sky background distribution is very important in designing a fiber spectroscopic survey with large field of view such as Large sky Area Multi-Object fiber Spectroscopic Telescope (LAMOST a.k.a. GuoShouJing telescope, \cite{2012RAA..12.1197C}). 
 
There are several sources that contribute to the night sky light after astronomical twilight, namely airglow,  scattering of starlight and zodiacal light, natural and artificial light pollution. For an astronomical observatory the artificial light pollution is generally very small and the most dominative light pollution source is the moonlight when the moon is above the horizontal. Many previous work has been done to study the night sky brightness, for example:\cite{skyb1},\cite{skyb2},\cite{skyb3}. However, most of the papers focus on the dark night
sky brightness, only  a few of them study the sky brightness in lunar nights. 
For LAMOST site,  the Xinglong Station of the National Astronomical Observatories in China (NAOC), which is located 100 km northwest to Beijing at a longitude of 7h50m18s east, a latitude of $40^{\circ}23'36''$ north, and an altitude of 950 m, 
 \cite{2003PASP..115..495L} and \cite{1674-4527-12-7-005} had studied the night sky brightness
, seeing, extinction, and available  observational hours as well as their seasonal changes   
  using photometric data of the BATC  telescope. For the moon night sky,  they use simple empirical model considering the moon phase, height and angular distance between the moon and target sky to eliminate the effect of the moon. The model
  used in their paper is too simple to estimate the detail sky brightness distribution in a 5 degree  field of view ( as LAMOST).  \cite{1991PASP..103.1033K} derived a more complicated model considering both the Rayleigh and Mie scattering of the moon light, two more variables, zenith distance of the moon and zenith distance of the sky were considered in their model,  the accuracy reported in their paper was 8-23\%.
 In the present work, with a modified moon night sky brightness model of \cite{1991PASP..103.1033K} and data from the measurement of a Sky Quality Meter (SQM) in Xinglong station, we derived the typical value for several important parameters in the model. With these results, we calculated the sky brightness distribution under a given moon condition. Furthermore, we use these results to predict the typical brightness gradient for LAMOST telescope 5 degree field of view.

In section {\ref{sect:model}} we describe the sky brightness model. In section {\ref{sect:sqmdata}} the process about how we derive the brightness measurement data from the SQM is explained. In section {\ref{sect:fitting}} we present the fitting of parameters, including the dark zenith sky brightness $V_{zen}$, the extinction coefficient k, the scattering coefficient PA and PB. In section {\ref{sect:calbrightness}} we calculate the sky brightness distribution when the moon is at a specific position and phase angle, using the typical value of deduced sky parameter . Furthermore the brightness variation distribution in a 5 degree field of view caused by the moon is predicted and the affection with the LAMOST is discussed.

\section{Calculation Model}           
\label{sect:model}

According to  \cite{1991PASP..103.1033K}, The sky brightness in lunar night can be divided into two part, the dark time night sky:
	
\begin{equation}
B_0(B_{zen},k,z)=B_{zen}10^{-0.4k(X-1)}X
\label{B0}
\end{equation}

 and the contribution from the scattered moon light:
 \begin{equation}
B_{moon}=f(\rho)I^*10^{-0.4kX_m(Z_m)}(1-10^{-0.4kX(Z)})
\label{BM}
\end{equation}

where $B_{zen}$ is the dark  time sky brightness at zenith in nanoLamberts (nL) which can be converted to  magnitude  $V_{zen} $ with equation (27) in \cite{1989PASP..101..306G} , k is the extinction coefficient, f($\rho$) is the scattering function at scattering angle $\rho$, $I^*$ is the brightness of the moon outside the atmosphere, can be expressed as a function of  moon phase angle $\alpha$ :
\begin{equation}
I^*=10^{-0.4(3.84+0.026|\alpha|+4\times10^{-9}{\alpha}^4)}
\end{equation}
Z is the zenith distance, X is the airmass, $X_m(Z_m)$ and X(Z) denote the airmass for the moon and the target sky respectively. The airmass X can be expressed as in \cite{1989PASP..101..306G} :
\begin{equation}
X(Z)=(1-0.96\sin^2{Z})^{-0.5}
\end{equation}

The scattering function is composed of two types of scattering in the atmosphere, the Rayleigh scattering from atmospheric gases and the Mie scattering by atmospheric aerosols. The Mie scattering may degenerate into Rayleigh scattering when the size of atmospheric aerosols decreases. Instead, when the size increase the scattering turn out to be geometrical optics. The scattering function is proportional to the fraction of incident light scattered into a unit solid angle with a scattering angle, and also varies with different wavelength.  According to \cite{1991PASP..103.1033K}   and \cite{2005BASI...33..513C}), we define the scattering functions as
\begin{equation}
f_R(\rho)=10^{5.36}(1.06+\cos^2{\rho})
\label{eq9}
\end{equation}
\begin{equation}
f_M(\rho)=10^{6.15-\frac{\rho}{40}}{, when }(\rho\ge10)
\end{equation}
\begin{equation}
f_M(\rho)=6.2\times10^7{\rho}^{-2}{, when }(\rho\le10)
\end{equation}
\begin{equation}
f(\rho)=PA{\times}f_M(\rho)+PB{\times}f_R(\rho)
\label{eq10}
\end{equation}
In these functions, $f_R(\rho)$ and $f_M(\rho)$ are  the Rayleigh and Mie scattering functions respectively.  $\rho$ is the scattering angle defined as the angular  separation between the moon and the sky position.  In equation \ref{eq10}, PA is the Mie scattering scale factor and PB is the Rayleigh scattering scale  factor,  they are proportional to the density of the Mie and Rayleigh scattering particles in the atmosphere. In \cite{1991PASP..103.1033K}, the scattering function(equation 16 in that paper) have absorbed constant factor relating to unit conversions and normalizations. Since the particle densities may change with time and weather conditions from site to site, we use PA and PB as scale factors to scale the relative particle density to the site (Mauna Kea) in that paper, 
and they were treated as  free parameters rather than  fixed values in the data fitting program in section \ref{sect:fitting} to deal with different weather conditions. If the scattering angle is small, the moonlight may directly enter the measurement instrument and equation \ref{eq9} is not applicative in that case.

Finally, the sky brightness in the moon night can be simply expressed as
\begin{equation}
B=B_0+B_{moon}
\end{equation}
Here, $B_0$ is the dark night sky brightness in equation \ref{B0} and $B_{moon}$ is the brightness caused by the moonlight in equation \ref{BM}. Combining all the equations before, the sky brightness can be expressed as
\begin{equation}
B=B(B_{zen},k,Z_{moon},Z_{sky},\rho,\alpha,PA,PB)
\label{eqn10}
\end{equation}
Here $B_{zen}$ is the dark zenith sky brightness in nL; k is the extinction coefficient; $Z_{moon}$ is the moon zenith distance; $Z_{sky}$ is the  zenith distance of sky position; $\rho$ is the scattering angle; $\alpha$ is the moon phase angle; PA is the Mie scattering scale factor; PB is the Rayleigh scattering scale factor.

The moon longitude and latitude can be calculated using Chapront ELP-2000/82, (\cite{Meeus:1991:AA:532892}). The equator coordinates (RA, DEC) can then be obtained. For application in this paper, the accuracy of the mean position of the moon is enough. The Greenwich Mean Sidereal Time can be calculated as
\begin{equation}
GMST=100.46061837+36000.770053608T+0.000387933T^2-T^3/38710000
\end{equation}
Here T is the Julian centuries from J2000.0. Then the hour angle of the moon can be calculated as
\begin{equation}
H_{local}=GMST-RA-LON
\end{equation}
Here the geographic longitude (LON) is $117.575703^\circ$ and the geographic latitude (LAT) is $40.3933333^\circ$. The Azimuth angle (A) and the elevation angle (h) of the moon can then be calculated as
\begin{equation}
\left\{
\begin{array}{c}
\sin{h}=\sin{LAT}\sin{DEC}+\cos{LAT}\cos{DEC}\cos{H_{local}} \\
\cos{h}cos{A}=\cos{LAT}\sin{DEC}-\sin{LAT}\cos{DEC}\cos{H_{local}} \\
\cos{h}sin{A}=-\cos{DEC}\sin{H_{local}}
\end{array}
\right.
\end{equation}

\section{Sky brightness data in Xinglong}
\label{sect:sqmdata}

Sky Quality Meter (SQM, see http://www.unihedron.com/projects/darksky/) is a handy tool
developed by Unihedron company to measure the sky brightness of visual light in 
$mag/arcsec^2$. There is a near-infrared blocking fiber to contain the light to visual band, the
transmission curve of SQM can be found in Fig22 of \cite{Cinzano2005}(http://www.lightpollution.it/download/sqmreport.pdf), the wavelength response is very broad, with half maximum of sensitive curve from 4000 to 6000\AA  \ and peaked at 5400\AA.  Due to the broadness of the response, the conversion to Johnson V band will depend on the spectral type.
But for dark night  or moon light sky brightness, the SQM magnitude is similar to V band, with an error of about 0.1 mag(\cite{Cinzano2005}). Each SQM is calibrated by a NIST-traceable light meter, the absolute precision of each meter is believed to be 10\%  ( 0.1$mag/ascsec^2$). The difference in zero point between each calibrated SQM is also 10\% (0.1$mag/ascsec^2$ ). Add the above errors together, the overall sky brightness measured in V band is about 0.2 $mag/arcsec^2$. 
  
A SQM is installed in the Xinglong station of NAOC. It measured the sky brightness in magnitude per square arcsecond per 6 minutes for more than a year. The Full Width Half Maximum (FWHM) of the angular sensitivity is $~20^\circ$. 
The SQM is fixed in a metal framework and will not change its direction until reinstalled   manually. Table \ref{tab1} shows the direction of SQM in degrees during the days when the data was produced.

\begin{figure}
\centering
\includegraphics[width=1.0\textwidth]{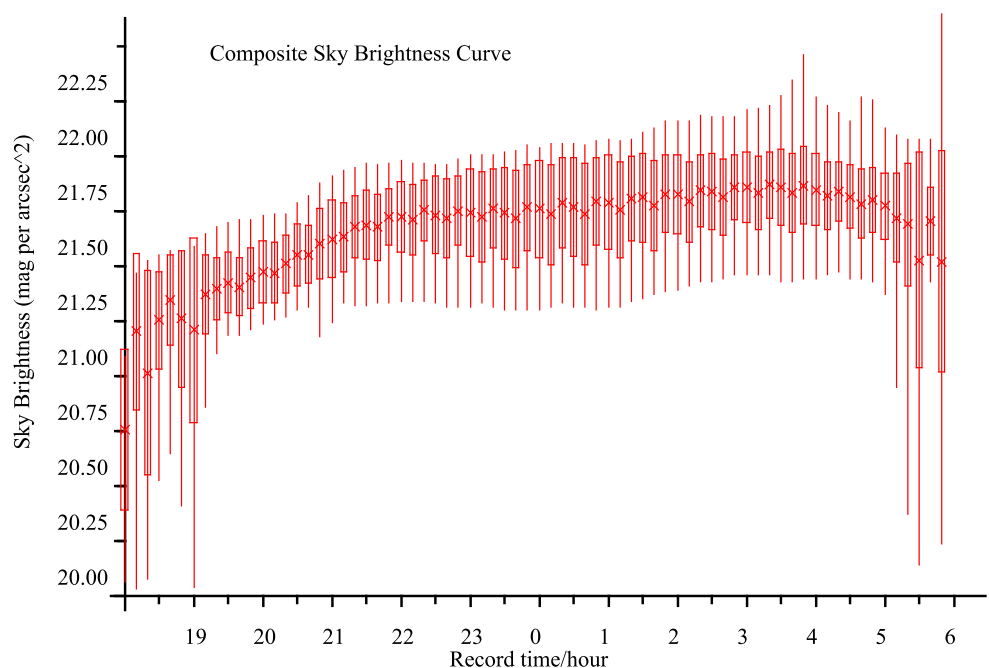}
\caption{The average of sky brightness with night time from 22 possible clear dark nights. The vertical box at each data point shows 1 $\sigma$ error bar,  vertical solid lines indicate the maximum and minimum value of that data point. There is a clear tendency that the night sky brightness is brighter in the first half of night, it goes darker in the second half due to less human activity . 
}
\label{fig:bvst}
\end{figure}

\begin{table}[h]
\begin{center}
\bc
\begin{minipage}[]{100mm}
\caption[]{Direction of the SQM\label{tab1}}\end{minipage}
\setlength{\tabcolsep}{1pt}
\small
 \begin{tabular}{cccc}
  \hline\noalign{\smallskip}
Start Date& End Date&  Azimuth($^\circ$)  &  Zenith Distance($^\circ$)\\
  \hline\noalign{\smallskip}
2010-12-16& 2011-12-14&1 80(South )&15\\
2011-12-15& 2012-01-12& 180(South) &30\\
2012-01-13& 2012-02-13& 90(East) & 30\\
2012-02-14& 2012-03-12& 0(North) & 30\\
  \noalign{\smallskip}\hline\\
  
\end{tabular}
\ec
\tablecomments{0.86\textwidth}{The azimuth zero point is the north and the clockwise direction is positive.}
\end{center}
\end{table}
To measure the sky brightness of dark nights, we pick out  data in the nights around new moon  when the moon is  below horizon, then we look up the observation log of LAMOST to reject those nights that was not marked  as clear night, this give us about 22 nights. It will be  interesting to explain how those data was influenced  by  light pollution from nearby cities and solar activities,  but as we only have a handful nights for each direction in 
table \ref{tab1},  and most data are in the winter, it's hard to tell how the artificial light influence the detail pattern of dark sky brightness. We still need  time to accumulate more data to find the nightly pattern with seasons and directions.
So, here we only show the average dark night brightness with time. We first correct the sky brightness to the zenith with equation\ref{B0}, then we set midnight as fiducial time, 
 the data from different nights were averaged every 10 minutes from the fiducial. The results
 is plotted in figure \ref{fig:bvst}, there is a clear tendency that the sky brightness is brighter in
 the first half of the night, after midnight the sky brightness gets darker about 0.3 magnitude.
 This may be explained as the human activities decrease after midnight, then the light pollution 
 is reduced at second half of night. The overall dark night sky brightness is about $21.6 \pm 0.2 mag/arcsec^2$. The error bar for each data point in figure \ref{fig:bvst} should reflect both the stability of the instrument and the nightly change of the sky condition, comparing with the sky brightness scatter in order of 0.4 magnitude in \cite{2003PASP..115..495L} paper,  also considering that our data were obtained in a period of about one year, the dominating contributor to this error bar should be the sky itself, the stability of SQM should be much smaller than 0.2 mag. 

The direction of the SQM was adjusted with simple tools so the accuracy is in several degrees. Those data were taken in more than 400 days with all kinds of weather conditions such as heavy cloudy days£¬ also there is no guarantee that the weather should be stable within one night. To measure the sky brightness of scattered  moonlight, we need to exclude those situation which the scattering model can not work properly , we use several rules to filter the proper data set:
\begin{enumerate}
\item Exclude the daytime records according to the astronomical twilight.
\item Exclude the records when the altitude of the moon is less than 5$^\circ$.
\item Exclude the records when the angular distance is smaller than 32.5$^\circ$, or the sky brightness is brighter than 18($mag/arcsec^2$) to avoid direct incidence of moon light.
\item Exclude the observation night whose record count is smaller than 25 to avoid  night with heavy cloud.
\item Exclude the observation night whose observed brightness curves show irregular transition or large dispersion than 30\% to avoid  drastic weather change or partial cloud blocking.
\end{enumerate}
In the end, we obtained sky brightness data for 114 moon nights from Dec 16th 2010 to Mar 12th 2012. Each observation night contains about 42 records.

\section{Parameters Fitting}
\label{sect:fitting}

This following work is mainly based on equation \ref{eqn10}. Here the zenith distance of the moon $Z_{moon}$, the zenith distance of the sky position $Z_{sky}$, the angular separation between the moon and the sky position $\rho$, and the moon phase angle $\alpha$  in degree are managed as input arguments, since they can be calculated from the observatory geographical location and time. The dark zenith sky brightness $V_{zen}$, the extinction coefficient k, the Mie scattering scale factor PA and the Rayleigh scattering scale factor PB are treated as parameters to be determined. The output result is sky brightness in nanoLamberts, which is provided through the SQM measurements.

In order to solve the non-linear least squares problem, we collect the observed sky brightness records on each observation night as one data set. The number of data records (N) in each data set is approximately 42. Using the observatory geographical location and time information, the moon and sky position related input values can be estimated and attached to each record. It is assumed that for one night the parameters ($V_{zen}$, k, PA and PB) do not change very much so they are treated as constants. We can determine the value of these parameters for each night that will give a minimum value of the squared 2-norm residual, which is defined as
\begin{equation}
RESNORM=\sum_{i=1}^{i=N}{(ModelB_i-ObservedB_i)^2}
\end{equation}
The initial value of the parameters ($V_{zen}$, k, PA and PB) are set as 21.4, 0.23, 1 and 1 respectively. The zenith magnitude is limited between 0 and 25; the extinction coefficient is limited between 0.01 and 8; the parameter PA and PB are limited between 0 and 25. Finally we obtained the fitting results for 114 data sets. If we reject data sets with returned parameters at or close to the limitation boundary, 90 data sets remained.

\begin{figure}
\centering
\includegraphics[width=0.60\textwidth]{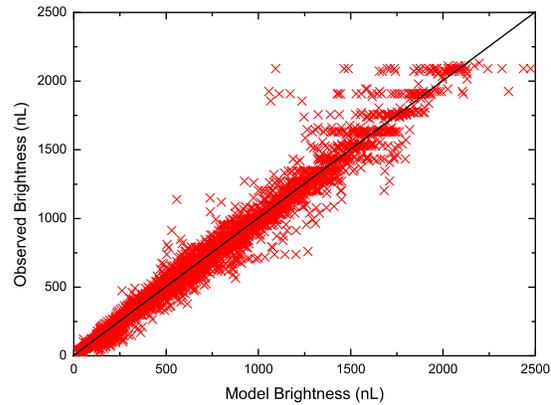}
\caption{The model brightness against the observed brightness. About 5200 data records from the 114 nights are presented. The total relative fitting variation defined by equation~(\ref{eq18}) is 12\%.
}
\label{fig:bvsb}
\end{figure}

\begin{figure}[h]
\centering
\includegraphics[width=0.60\textwidth]{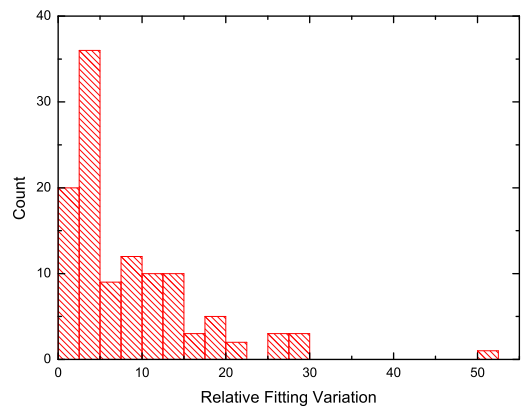}
\caption{Histogram of relative fitting variation of each observation night.
Only the the last remaining 90 nights are presented. Most of the variations are smaller than 5\%.
}
\label{fig:resnorm}
\end{figure}

Figure~\ref{fig:bvsb} shows the model brightness against the measured brightness for the data records of all the data sets from 114 nights, containing more than 5000 records. We define the Relative Fitting Variation (RFV) as
\begin{equation}
RFV=
\sqrt{\frac{1}{N}\sum_{i=1}^{i=N}\left(\frac
{
ModelB_i-ObservedB_i
}
{
{ObservedB_i}
}\right)^2}
\label{eq18}
\end{equation}
The total relative fitting variation defined by equation~(\ref{eq18}) in Figure~\ref{fig:bvsb} is 12\%. The RFVs for each data set are shown in Figure~\ref{fig:resnorm}. Only the last remaining 90 data sets are presented. As we can see that most of fitting results have a RFV smaller than 5\%. That means for most of the remaining nights, the model brightness matches the observed brightness with a nice accuracy.

\begin{figure}[h]
\begin{center}
\includegraphics[width=0.90\textwidth, height=0.75\textwidth]{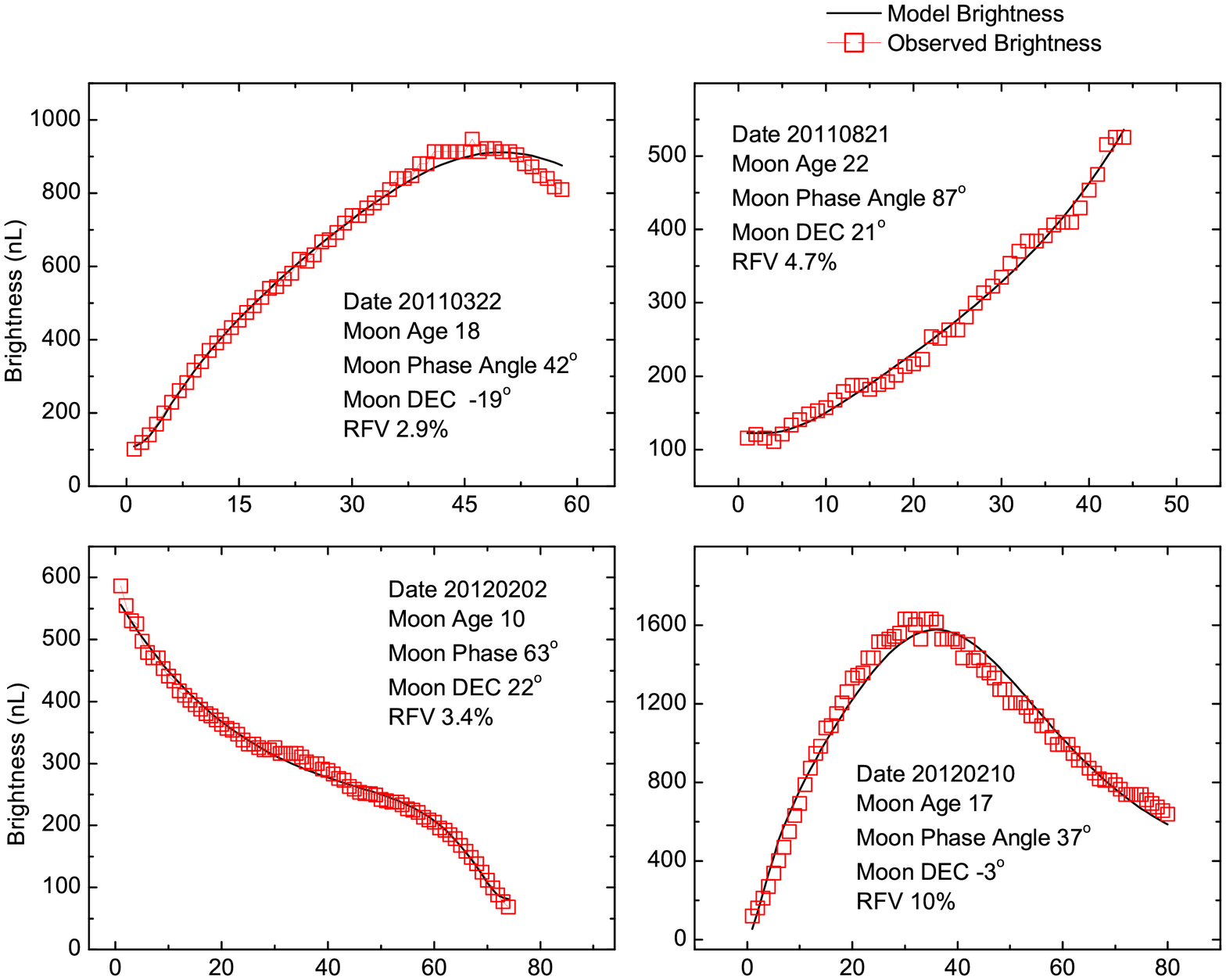}
\caption{Sky brightness fitting curves. The X axis is the serial number for the measured data point. Related parameters, including observation date, moon age, moon phase angle in degrees, moon declination and relative fitting variation RFV, are printed in the figures. The definition of RFV is described in equation~(\ref{eq18}). The SQM directions are listed in table \ref{tab1}.
}
\label{fig:fitcurve}
\end{center}
\end{figure}

Figure~\ref{fig:fitcurve} shows  examples of fitting curve  from four nights. The moon phase angle here is expressed as the Sun-Moon-Earth angle in degrees. The relative fitting variations are 2.9\%, 4.7\%, 3.5\% and 10\% respectively, as marked in the figure. It is obvious that the bottom-right figure has the maximum brightness because the phase angle is the smallest among the four figures, while the top-right one has the minimum brightness. The major influence factor of the curve trend is the moon/sky angular separation. As we can see from these figures that when the moon moves close to the sky position, the sky background becomes bright. Since the sky direction and the moon declination for one night is basically fixed, the angular separation is determined by the moon altitude. Eventually when the moon is on the meridian, the background in the field of view of a quasi-meridian telescope, such as LAMOST, will be very bright. The results shows that this fitting model works well for various phase angle and location of the moon.

Nights with bad fitting results are due to following reasons:
\begin{enumerate}
\item The weather condition. As this measurement instrument keeps working without interruption for years, the data is produced on days under various kinds of weather condition. These affection cannot be corrected using the presented models.
\item The large angular sensitivity of the SQM. Within that large field of view the light from stars and the planets, especially the moonlight when the moon is close, will greatly affect the measure results. As a result, the measured sky brightness shows a sharp increase as the moon moves close to the sky position.
\item The reflection of the moonlight from the surrounding buildings.
\item The light pollution from the nearby cities. The sky brightness of Xinglong station is mainly polluted by the city lights from Beijing, Xinglong, and Chengde. As we can see from figure{\ref{fig:bvst}}, the artificial light pollution may change the sky brightness about 0.3 mag at night. Those affections are ignored in the model since the sky brightness we are considering is at least one magnitude brighter than the dark night. The sky brightness in the absence of moonlight is expressed as equation~(\ref{B0}).
\end{enumerate}

\begin{figure}
\centering
\includegraphics[width=0.90\textwidth]{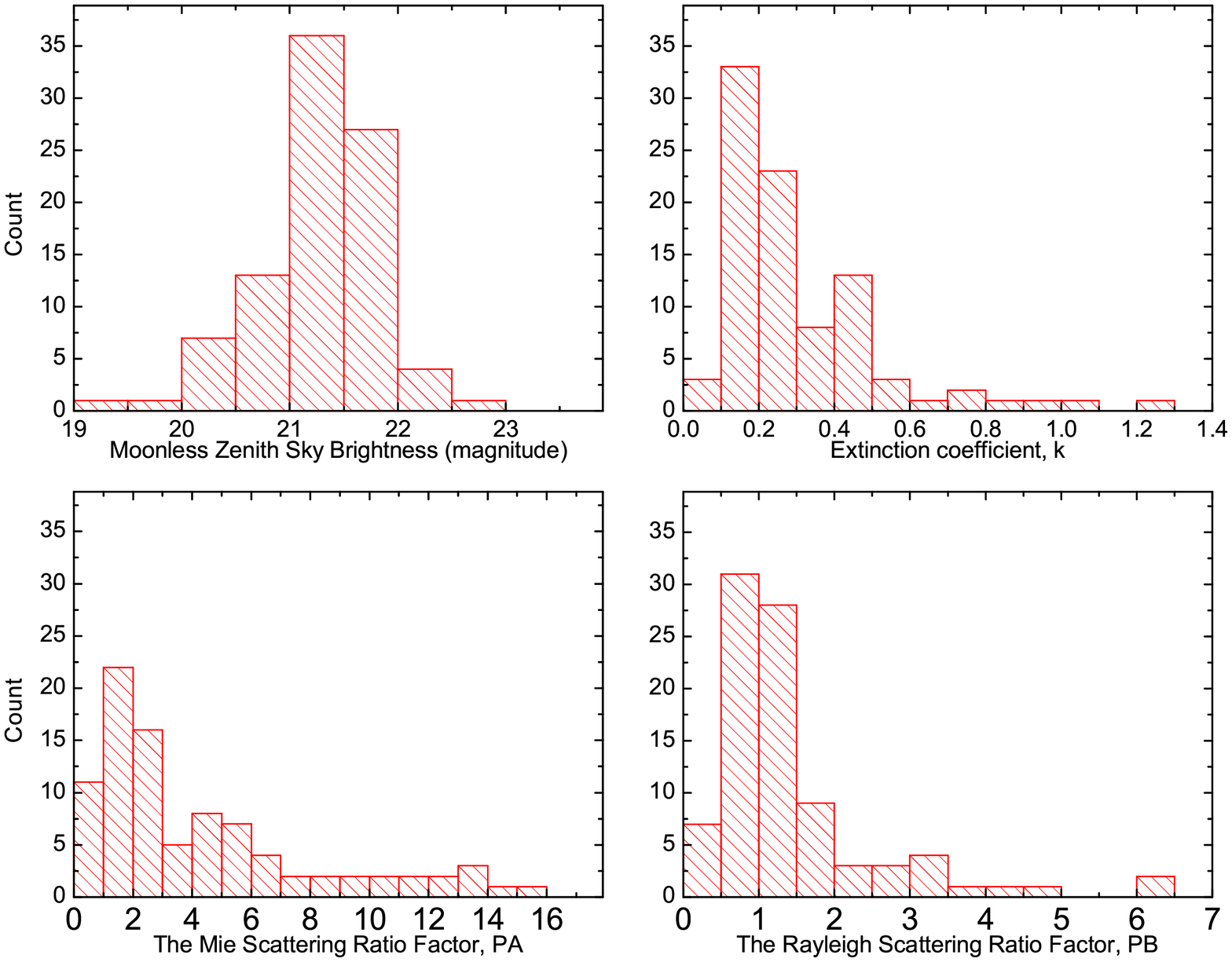}
\caption{Histogram of the estimated parameters including the dark zenith sky brightness $V_{zen}$, extinction coefficient k, scattering coefficients PA and PB. Only the last remaining 90 data sets are presented.  The typical value for these parameters can be read from the figures as:  $V_{zen}$=21.4, k=0.23, PA=1.5 and PB=0.90}
\label{fig:hist}
\end{figure}

After iterative fitting, the result histogram of the estimated parameters including the dark zenith sky brightness $V_{zen}$, extinction coefficient k, scattering coefficients PA and PB are shown in Figure~\ref{fig:hist}. Only the last remaining 90 data sets are presented. The dark zenith sky brightness, $V_{zen}$, presents a typical value of 21.4 $mag/arcsec^2$. This result agrees with the zenith sky brightness ($21.6 \pm 0.2 mag/arcsec^2$) in the winter we measured in section   \ref{sect:sqmdata} from dark nights. This also basically fit the sky brightness in the V band obtained from BATC Polaris monitoring data (\cite{1674-4527-12-7-005}), which is about $21 mag/arcsec^2$ at the North Pole. Correcting the airmass using equation \ref{B0} for 0.36 magnitude, the results agree very well. The extinction coefficient, k, presents a typical value of 0.23. This is in the acceptable range according to the measurement results in Xinglong (\cite{2003PASP..115..495L}; \cite{1674-4527-12-7-005}). The Mie scattering scale factor, PA, presents a typical value of 1.50. The Rayleigh scattering scale factor, PB, presents a typical value of 0.90.

Figure~\ref{fig:hist} shows the estimated parameters as a function of date from January 1 2011.  
The data were collected within one year,   there is an obvious  lack of data in the summer due the bad weather. Since there are not enough data, it's hard to tell the system change of those parameters with seasons. However we can still find some indication that the dark zenith sky is darker in the winter than in the summer and the sky transparency in the autumn and  winter  is better than those in the spring  and summer, which is also consistent with previous work, e.g. \cite{2003PASP..115..495L} and \cite{1674-4527-12-7-005}.  While the scattering scale factor PA and PB show no obvious seasonal variation. The factor PA has a larger dispersion than  PB. This is mainly because the density of the Mie scattering particles (i.e. dust) has a bigger variation than the Rayleigh molecule. More data are need to better resolve the seasonal change of these parameters.

\begin{figure}
\centering
\includegraphics[width=0.90\textwidth]{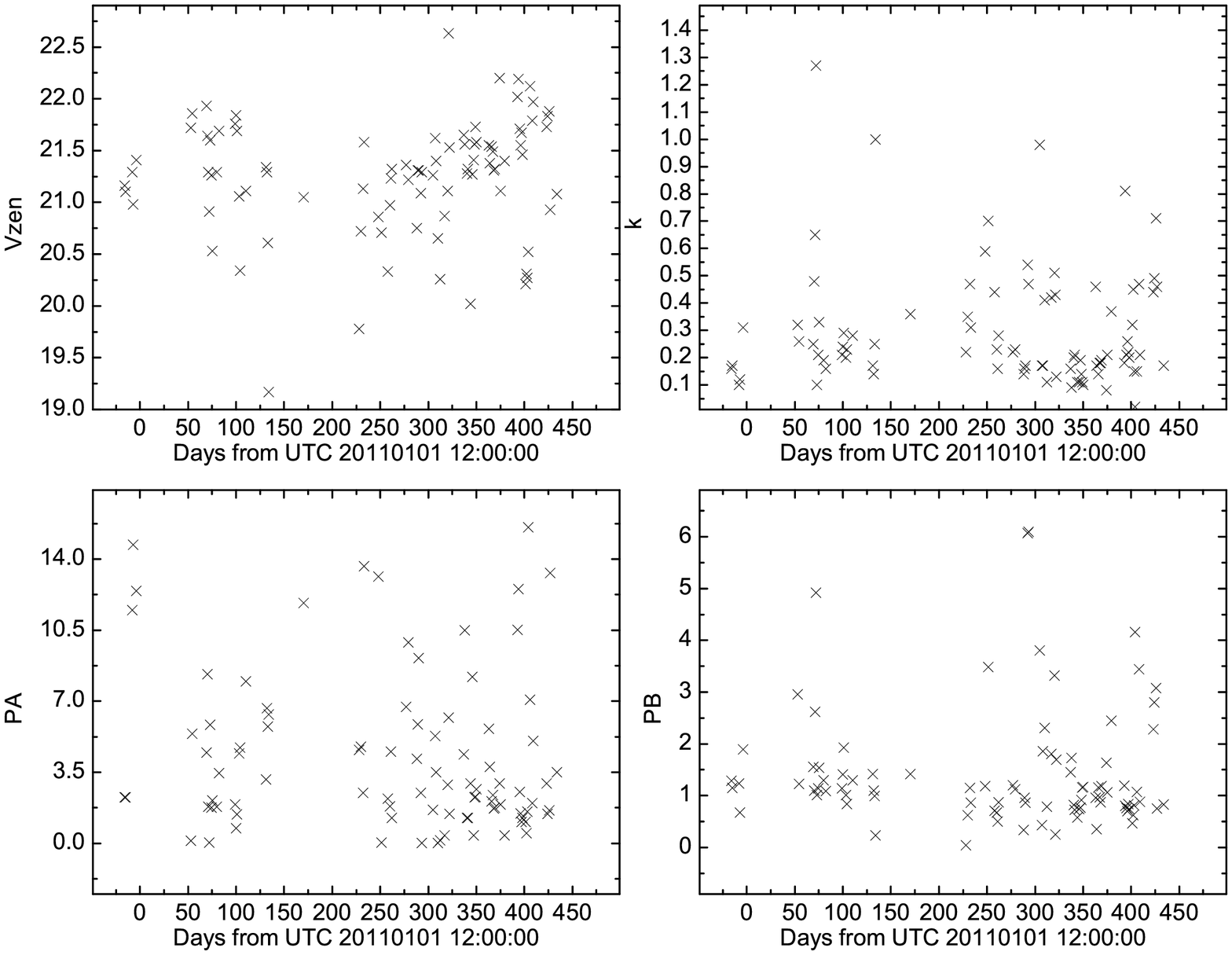}
\caption{Estimated parameters on days. Top left: zenith sky brightness vs. time. Top right: extinction vs. time. Bottom left: Mie scattering factor PA vs. time. Bottom right: Rayleigh scattering factor PB vs. time.  Only the last remaining 90 data sets are presented. The days are calculated from 1 January 2011, 12:00:00. Days close to 0 or 365.25 are in the winter. Days near 180 are in the summer. }
\label{fig:ondays}
\end{figure}
\begin{figure}
\centering
\includegraphics[width=0.7\textwidth]{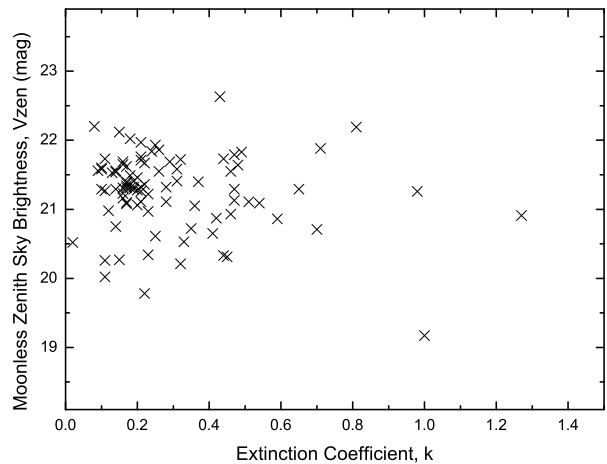}
\caption{ extinction coefficient k vs. zenith sky brightness$V_{zen}$}
\label{fig:VzenK}
\end{figure}

In \cite{2003PASP..115..495L} paper, they found a  linear relation between the sky brightness and extinction coefficient k. While in our results( see figure \ref{fig:VzenK}), the correlation is not found. Since the data in 
\cite{2003PASP..115..495L} paper were collected around Polaris, there is a fix zenith distance about $50^\circ$
in Xinglong station, considering equation \ref{B0}, we conclude that the linear correlation agrees with the equation while the measured sky brightness is NOT the zenith brightness.

\FloatBarrier
\section{Sky brightness distribution}
\label{sect:calbrightness}

\subsection{Sky brightness estimation}
\label{sec:subsec5}
\begin{figure}
\centering
\includegraphics[width=1.00\textwidth]{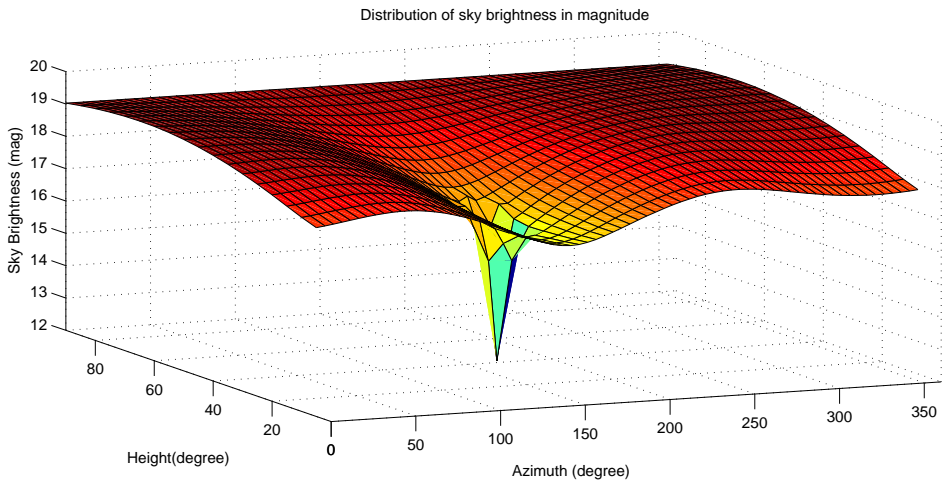}
\caption{Distribution of sky brightness in magnitude, the input parameters are: $V_{zen}$=21.4, K=0.23, Moon phase=$20^{\circ}$, DEC=$-20^{\circ}$, hour angle=$-30^{\circ}$, PA=1.5, PB=0.90 .}
\label{fig:sky}
\end{figure}

Estimation of sky background is important for a telescope working at lunar night. The background brightness is one of the
key factor to determine the limiting magnitude and exposure time. For a survey telescope working at lunar night, knowing the background distribution will help astronomers to design how faraway from the moon should the telescope point to  in order to reduce the influence of the moon. By applying the typical parameters of fitting results to equation~(\ref{eqn10}),  we can calculate the sky brightness for Xinglong station given sky position and time  within lunar night, except the region where the moon is very close to the sky position ($< 10^\circ$).  Here  we give two examples in table~\ref{s10before} and table~\ref{phase90}. 
In table~\ref{s10before} the moon is assumed to be with a phase angle of 20$^\circ$, an hour angle of -30$^\circ$ and a declination of -10$^\circ$. The brightness distribution  in a large region with hour angle from -35$^\circ$ to 35$^\circ$ and declination from -10$^\circ$ to 70$^\circ$ is presented. For an intuitive view,  sky brightness distribution for moon phase angle $20^{\circ}$,  declination $-20^{\circ}$ and hour angle $-30^{\circ}$ is plotted in figure \ref{fig:sky}. Table \ref{phase90} is for a moon on the meridian with phase angle $90^{\circ}$, declination $20^{\circ}$. The position within $10^\circ$ of the moon was left blank in the table. The brightness in table \ref{s10before} and \ref{phase90} will be used to set the current magnitude limit and estimate exposure time  for LAMOST survey in each lunar night. They could also help to determine  where the telescope should point to given the scientific request. For example, to reach magnitude 17  in bright night, the background should be no brighter than 19$mag/arcsec^2$.  In table \ref{s10before} and figure \ref{fig:sky}, the moon is close to full moon, sky should be at least  $65^\circ$ away to reach 19 $mag/arcsec^2$, comparing with table \ref{phase90} for which the moon is at half moon, the sky is as deep as 19 $mag/arcsec^2$ when $15^\circ$ from the moon. From table \ref{s10before}, \ref{phase90}  and figure \ref{fig:sky}, we can see  the brightness gradient is larger when closer to the moon. The sky brightness drops faster in declination directions than hour angle direction. In declination direction, sky brightness drops faster  when pointing to the zenith than to the horizon. 

\begin{table}
\bc
\begin{minipage}[]{130mm}
\caption[]{The sky brightness distribution in Xinglong station when the moon is 2 hours before the transit. The brightness
is in V band $mag/arcsec^2$. input conditions are listed below the table. \label{s10before}}\end{minipage}
\setlength{\tabcolsep}{1pt}
\small
 \begin{tabular}{cccccccccccccccc}
  \hline\noalign{\smallskip}
Dec$\backslash$h&-35&-30&-25&-20&-15&-10&-5&0&5&10&15&20&25&30&35\\
  \hline\noalign{\smallskip}\\
-10&  	&Moon&		&	&17.16&17.39&17.59&17.77&17.93&18.08&18.20&18.30&18.38&18.45&18.49\\
-5 &	&	&	&17.06&17.29&17.50&17.70&17.88&18.03&18.17&18.29&18.40&18.48&18.55&18.60\\
0  &17.02&	&17.10&17.27&17.45&17.64&17.83&17.99&18.14&18.28&18.39&18.49&18.58&18.64&18.69\\
5  &17.29&17.30&17.37&17.50&17.65&17.81&17.97&18.12&18.26&18.38&18.49&18.59&18.67&18.73&18.78\\
10 &17.55&17.57&17.63&17.72&17.84&17.98&18.11&18.25&18.37&18.49&18.59&18.68&18.75&18.81&18.86\\
15 &17.79&17.81&17.86&17.94&18.04&18.15&18.26&18.38&18.49&18.59&18.68&18.76&18.83&18.89&18.93\\
20 &18.01&18.03&18.07&18.14&18.22&18.31&18.41&18.51&18.60&18.69&18.77&18.85&18.91&18.96&18.99\\
25 &18.20&18.23&18.27&18.32&18.39&18.46&18.54&18.63&18.71&18.79&18.86&18.92&18.98&19.02&19.05\\
30 &18.38&18.40&18.44&18.48&18.54&18.60&18.67&18.74&18.81&18.88&18.94&18.99&19.04&19.08&19.10\\
35 &18.54&18.56&18.59&18.63&18.68&18.73&18.79&18.85&18.90&18.96&19.01&19.06&19.10&19.13&19.14\\
40 &18.68&18.70&18.73&18.76&18.80&18.84&18.89&18.94&18.99&19.03&19.07&19.11&19.14&19.17&19.18\\
45 &18.80&18.82&18.84&18.87&18.91&18.94&18.98&19.02&19.06&19.10&19.13&19.16&19.18&19.20&19.20\\
50 &18.90&18.92&18.94&18.97&19.00&19.03&19.06&19.09&19.12&19.15&19.17&19.19&19.21&19.22&19.22\\
55 &18.99&19.01&19.03&19.05&19.07&19.09&19.12&19.14&19.16&19.18&19.20&19.22&19.23&19.23&19.23\\
60 &19.05&19.07&19.09&19.11&19.12&19.14&19.16&19.18&19.19&19.21&19.22&19.23&19.23&19.23&19.22\\
65 &19.10&19.11&19.13&19.14&19.16&19.17&19.18&19.20&19.21&19.21&19.22&19.22&19.22&19.22&19.21\\
70 &19.12&19.14&19.15&19.16&19.17&19.18&19.19&19.20&19.20&19.20&19.21&19.21&19.20&19.20&19.19\\
  \noalign{\smallskip}\hline
\end{tabular}
\ec
\tablecomments{0.86\textwidth}{Unit:mag. $V_{zen}$=21.4, k=0.23, moon phase=20$^\circ$, hour angle=-30$^\circ$, DEC=-10$^\circ$, PA=1.5, PB=0.90}
\end{table}

\begin{table}
\bc
\begin{minipage}[]{130mm}
\caption[]{The sky brightness distribution when moon phase angle is 90 degrees\label{phase90}}\end{minipage}
\setlength{\tabcolsep}{1pt}
\small
 \begin{tabular}{cccccccccccccccc}
  \hline\noalign{\smallskip}
Dec$\backslash$h&-35&-30&-25&-20&-15&-10&-5&0&5&10&15&20&25&30&35\\
  \hline\noalign{\smallskip}\\
-10&19.53&19.50&19.46&19.42&19.38&19.35&19.33&19.32&19.33&19.35&19.38&19.42&19.46&19.50&19.53\\
-5 &19.55&19.50&19.45&19.39&19.34&19.29&19.26&19.24&19.26&19.29&19.34&19.39&19.45&19.50&19.55\\
0  &19.56&19.50&19.43&19.35&19.28&19.21&19.16&19.14&19.16&19.21&19.28&19.35&19.43&19.50&19.56\\
5  &19.56&19.49&19.40&19.31&19.21&19.11&19.04&19.01&19.04&19.11&19.21&19.31&19.40&19.49&19.56\\
10 &19.57&19.48&19.38&19.26&19.14&19.01&18.90&	&18.90&19.01&19.14&19.26&19.38&19.48&19.57\\
15 &19.58&19.48&19.37&19.23&19.08&18.92&		&	&	&18.92&19.08&19.23&19.37&19.48&19.58\\
20 &19.59&19.49&19.38&19.24&19.08&	&		&Moon&	&	&19.08&19.24&19.38&19.49&19.59\\
25 &19.62&19.52&19.41&19.28&19.13&18.97&		&	&	&18.97&19.13&19.28&19.41&19.52&19.62\\
30 &19.66&19.57&19.47&19.35&19.23&19.11&19.02&	&19.02&19.11&19.23&19.35&19.47&19.57&19.66\\
35 &19.71&19.63&19.54&19.45&19.36&19.27&19.21&19.19&19.21&19.27&19.36&19.45&19.54&19.63&19.71\\
40 &19.76&19.70&19.63&19.56&19.49&19.43&19.39&19.38&19.39&19.43&19.49&19.56&19.63&19.70&19.76\\
45 &19.82&19.77&19.72&19.67&19.62&19.58&19.55&19.55&19.55&19.58&19.62&19.67&19.72&19.77&19.82\\
50 &19.88&19.84&19.81&19.77&19.74&19.71&19.69&19.69&19.69&19.71&19.74&19.77&19.81&19.84&19.88\\
55 &19.94&19.91&19.89&19.86&19.84&19.82&19.81&19.81&19.81&19.82&19.84&19.86&19.89&19.91&19.94\\
60 &19.99&19.97&19.95&19.94&19.92&19.91&19.91&19.90&19.91&19.91&19.92&19.94&19.95&19.97&19.99\\
65 &20.03&20.02&20.01&20.00&19.99&19.99&19.98&19.98&19.98&19.99&19.99&20.00&20.01&20.02&20.03\\
70 &20.07&20.06&20.06&20.05&20.05&20.04&20.04&20.04&20.04&20.04&20.05&20.05&20.06&20.06&20.07\\
  \noalign{\smallskip}\hline
\end{tabular}
\ec
\tablecomments{0.86\textwidth}{Unit:mag. $V_{zen}$=21.4, k=0.23, moon phase=90$^\circ$, hour angle=0$^\circ$, DEC=20$^\circ$, PA=1.5, PB=0.90}
\end{table}

\subsection{Brightness difference within a 5 degree field of view}
\label{sect:caldiff}

\begin{figure}[h]
\centering
\includegraphics[width=0.90\textwidth]{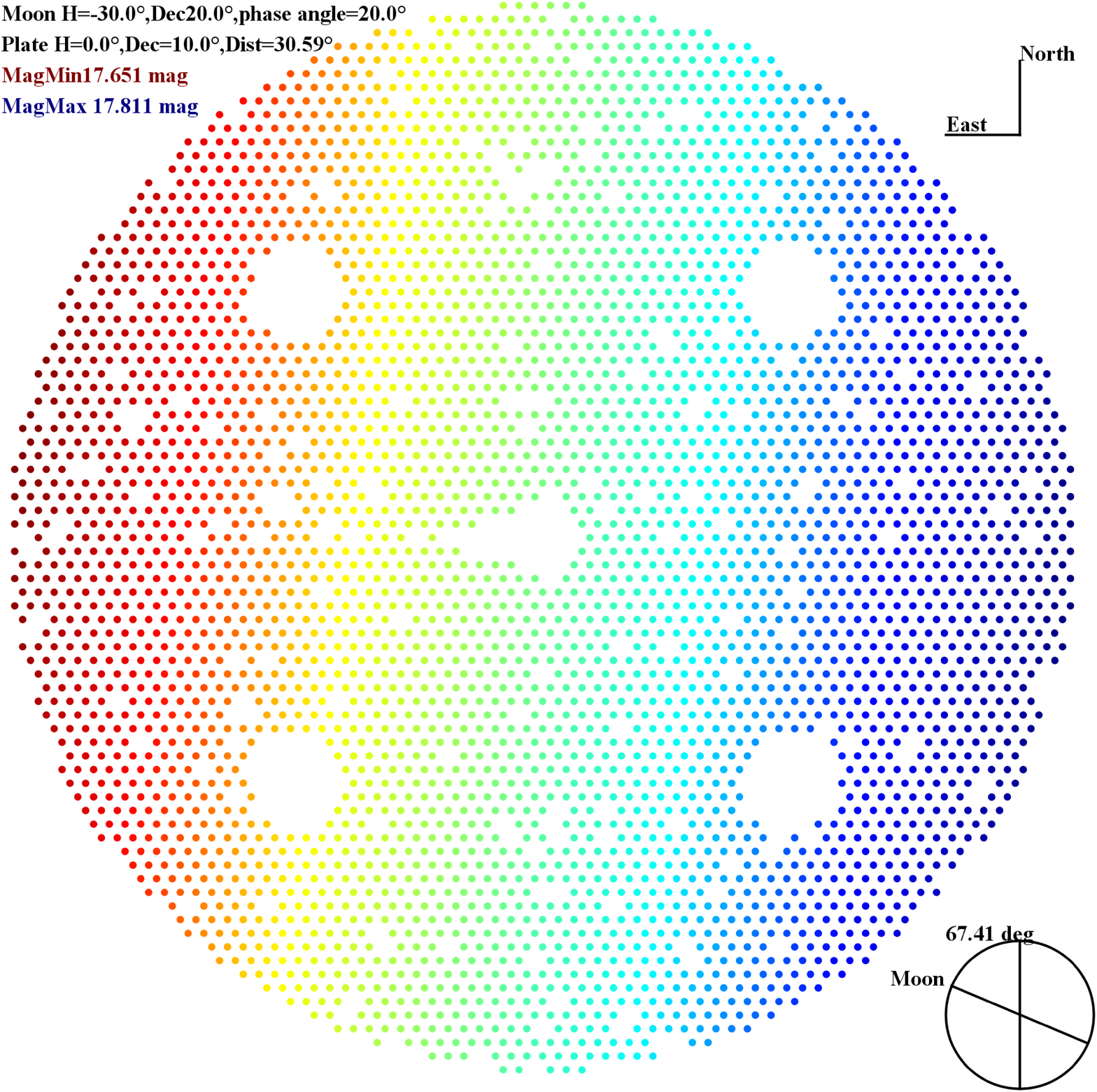}
\caption{
Simulation of brightness distribution in LAMOST focal plate. The moon is 2 hours before the zenith and the declination is 20$^\circ$. The telescope is pointing to transit with declination of 10$^\circ$. PA = 1.5, PB = 0.90, $V_{zen}$ = 21.4, k = 0.23, moon phase angle $\alpha$= 20$^\circ$. The fiber location is assumed to be placed in the sky according to their light path to make the figure more intuitive. Affected by the factor relating the zenith distance of the sky position, the moon direction does not match with the gradient direction of brightness. The increasing direction of brightness shifts from the moon to the horizon slightly. The variation becomes larger at the horizon.
}
\label{fig:focal}
\end{figure}

The Large Sky Area Multi-Object Fiber Spectroscopic Telescope (LAMOST) is an optical  survey telescope with a field of view of 5 degrees.  There are 16 spectrographs each holding 250 fibers, each spectrograph occupy a certain field of view( about  $1^\circ$) on the LAMOST focal plane. In such a large field of view, the  background gradient can not be ignored.  In  most  fiber spectra 
observations,  the sky is sampled by dedicated fibers, then the sky in the object fiber is subtracted using a  spectrum
 composited from those sky fibers. This step works well when the sky is homogenous within the field of view of spectrograph,
 in the case of moon night, the  sky background gradient  should be considered in the data reduction step as well as in the step that designing how the sky fiber   sampled the background. 
 In this work, we try to estimate the brightness difference caused by the moonlight within the field of view of LAMOST, so as to provide reference basis for observation strategy  decision as well as data reduction. From table \ref{s10before} and \ref{phase90}, it is easy to tell that the larger the angular separation, the less the sky brightness gradient is. To calculate the maximum brightness difference within the field of view, we need to find out the point with the maximum and  minimum brightness. 
The points with extreme value was found by a step by step search along the edge of field of view. We show an example of results  in table \ref{diff.phase.20}.   The input parameters
are the same as in table \ref{phase90} except that the moon phase angle is $20^\circ$ rather than $90^\circ$. 
Each item in table \ref{diff.phase.20} is the max difference in LAMOST 5 degree field of view. The max difference in
table \ref{diff.phase.20} is 0.36 $mag/5^\circ$,  so in the field of view of one LAMOST spectrograph (about $1^\circ$) , the gradient will be 
about 0.07 mag. Which means the gradient even in one spectrograph is 7\%, this will cause  sky subtraction residuals
in data reduction if sky is not properly sampled. While the smallest difference in the table is about 0.05 mag which means about 1\% gradient in one spectrograph. Generally, the sky subtraction accuracy in fiber spectra data reduction is larger than $2\%$,
1\% difference can be acceptable in most cases.  From the above discussion,   it's better to choose the position with 
smaller background gradient to  alleviate data reduction difficulty in designing a survey like LAMOST. 
  Figure \ref{fig:focal} shows an example of sky brightness distribution inside LAMOST 5 degree field of view,
we calculate the brightness for each fiber. The moon phase angle is $20^\circ$, the distance between the moon and the field center is about $30^\circ$ as marked in the figure. The moon is $67^\circ$ to the north east, as show by the lower right icon.
 As we can see, there is a brightness gradient  about 0.16 magnitude, but the gradient direction is not the direction from the moon to the sky position. As we pointed out in section \ref{sec:subsec5}, this is because when the airmass increase against the  zenith direction( i.e. north in this figure), the moon light gets more scattered by the atmosphere,  this will change the direction of gradient a bit to the south . 


\begin{table}
\bc
\begin{minipage}[]{100mm}
\caption[]{Brightness difference within LAMOST field of view when the moon is over the zenith\label{diff.phase.20}}\end{minipage}
\setlength{\tabcolsep}{1pt}
\small
 \begin{tabular}{cccccccccccccccc}
  \hline\noalign{\smallskip}
Dec$\backslash$h&-35&-30&-25&-20&-15&-10&-5&0&5&10&15&20&25&30&35\\
  \hline\noalign{\smallskip}\\
-10&0.05&0.06&0.06&0.07&0.08&0.09&0.09&0.09&0.09&0.09&0.08&0.07&0.06&0.06&0.05\\
-5 &0.06&0.07&0.08&0.10&0.10&0.11&0.11&0.11&0.11&0.11&0.10&0.10&0.08&0.07&0.06\\
0  &0.08&0.09&0.10&0.12&0.13&0.13&0.14&0.14&0.14&0.13&0.13&0.12&0.10&0.09&0.08\\
5  &0.09&0.11&0.12&0.14&0.15&0.16&0.16&0.16&0.16&0.16&0.15&0.14&0.12&0.11&0.09\\
10 &0.11&0.12&0.14&0.16&0.17&0.18&0.23&	&0.23&0.18&0.17&0.16&0.14&0.12&0.11\\
15 &0.12&0.14&0.15&0.17&0.19&0.29&	&	&	&0.29&0.19&0.17&0.15&0.14&0.12\\
20 &0.13&0.15&0.17&0.19&0.20&	&	&Moon & & &0.20&0.19&0.17&0.15&0.13\\
25 &0.13&0.15&0.17&0.19&0.21&0.36&	& 	&	&0.36&0.21&0.19&0.17&0.15&0.13\\
30 &0.13&0.15&0.17&0.19&0.21&0.23&0.26& 	&0.26&0.23&0.21&0.19&0.17&0.15&0.13\\
35 &0.13&0.15&0.17&0.18&0.20&0.21&0.22&0.23&0.22&0.21&0.20&0.18&0.17&0.15&0.13\\
40 &0.13&0.14&0.16&0.17&0.18&0.19&0.20&0.20&0.20&0.19&0.18&0.17&0.16&0.14&0.13\\
45 &0.12&0.14&0.15&0.16&0.17&0.18&0.18&0.18&0.18&0.18&0.17&0.16&0.15&0.14&0.12\\
50 &0.11&0.12&0.13&0.14&0.15&0.16&0.16&0.16&0.16&0.16&0.15&0.14&0.13&0.12&0.11\\
55 &0.10&0.11&0.12&0.13&0.13&0.14&0.14&0.14&0.14&0.14&0.13&0.13&0.12&0.11&0.10\\
60 &0.09&0.10&0.11&0.11&0.11&0.12&0.12&0.12&0.12&0.12&0.11&0.11&0.11&0.10&0.09\\
65 &0.08&0.08&0.09&0.09&0.10&0.10&0.10&0.10&0.10&0.10&0.10&0.09&0.09&0.08&0.08\\
70 &0.07&0.07&0.07&0.08&0.08&0.08&0.08&0.08&0.08&0.08&0.08&0.08&0.07&0.07&0.07\\
  \noalign{\smallskip}\hline
\end{tabular}
\ec
\tablecomments{0.86\textwidth}{Unit:mag. $V_{zen}$=21.4, k=0.23, moon phase=20$^\circ$, hour angle=0$^\circ$, DEC=20$^\circ$, PA=1.5, PB=0.90}
\end{table}

\section{Summaries}
\label{sect:Summaries}

We use a SQM to study the night sky brightness  in Xinglong station. From the collected data from Dec 2010 to Mar 2012,
we selected 22 dark clear nights to  study the sky brightness variation with time, we found a clear correlation between the dark night sky brightness with human activity. 
We also study the lunar sky brightness model of \cite{1991PASP..103.1033K}, by modifying the relative scale factors of Rayleigh and Mie scattering respectively, we successfully fitted  the sky brightness data of SQM in Xinglong station with  a relative fitting variation of 12\%. We estimate the related parameters in 90 nights. According to the results we can see that in this observatory the typical  V band dark zenith sky brightness is about 21.4 $mag/ascsec^2$; the extinction coefficient is about 0.23; the Mie scattering scale factor is about 1.5; the Rayleigh scattering scale factor is about 0.90.
With the model and those typical parameters for Xinglong station, we could estimate the sky brightness distribution for any given time in moon night for the LAMOST site.  We then  calculated the gradient within the LAMOST 5 degree field of view.  The result shows that sky brightness increases quickly as the distance of the moon is smaller. The increasing zenith distance will also enhance the brightness within a quantity much smaller than the influence of the lunar separation angle. These results will help in designing the LAMOST bright night survey, determine the location of sky fiber in the focal plane as well as  data reduction in LAMOST survey.

\normalem
\begin{acknowledgements}
The author thanks Qiu Peng  and Lu Xiaomeng for helping to get the SQM data and informations. The author also
thanks the referee for the useful suggestions.
The Guoshoujing Telescope (the Large Sky Area Multi-Object Fiber
Spectroscopic Telescope; LAMOST) is a National Major Scientific Project built by the Chinese
Academy of Sciences. Funding for the project has been provided by the National Development
and Reform Commission. The LAMOST is operated and managed by the National Astronomical
Observatories, Chinese Academy of Sciences.
\end{acknowledgements}

\bibliographystyle{raa}
\bibliography{bibtex}

\begin{thebibliography}{11}
\providecommand{\natexlab}[1]{#1}
\providecommand{\selectlanguage}[1]{\relax}

\bibitem[{{Chakraborty} et~al.(2005){Chakraborty}, {Das1}, \&
  {Tandon}}]{2005BASI...33..513C}
{Chakraborty}, P., {Das1}, H.~K., \& {Tandon}, S.~N. 2005, Bulletin of the
  Astronomical Society of India, 33, 513

\bibitem[{Cinzano(2005)}]{Cinzano2005}
Cinzano, P. 2005, ISTIL Int. Rep., 9,
  http://www.lightpollution.it/download/sqmreport.pdf

\bibitem[{{Cui} et~al.(2012){Cui}, {Zhao}, {Chu} et~al.}]{2012RAA..12.1197C}
{Cui}, X., {Zhao}, Y., {Chu}, Y., et~al. 2012, \raa, 12, 1197

\bibitem[{{Garstang}(1989)}]{1989PASP..101..306G}
{Garstang}, R.~H. 1989, \pasp, 101, 306

\bibitem[{{Krisciunas} \& {Schaefer}(1991)}]{1991PASP..103.1033K}
{Krisciunas}, K., \& {Schaefer}, B.~E. 1991, \pasp, 103, 1033

\bibitem[{{Liu} et~al.(2003){Liu}, {Zhou}, {Sun} et~al.}]{2003PASP..115..495L}
{Liu}, Y., {Zhou}, X., {Sun}, W.-H., et~al. 2003, \pasp, 115, 495

\bibitem[{Meeus(1991)}]{Meeus:1991:AA:532892}
Meeus, J.~H. 1991, Astronomical Algorithms (Willmann-Bell, Incorporated)

\bibitem[{{Neugent} \& {Massey}(2010)}]{skyb3}
{Neugent}, K.~F., \& {Massey}, P. 2010, \pasp, 122, 1246

\bibitem[{{Sanchez} et~al.(2007){Sanchez}, {Aceitune}, {Thiele}, \&
  {Alves}}]{skyb1}
{Sanchez}, S., {Aceitune}, J., {Thiele}, D., U.~and{Perez-Ramirez}, \& {Alves},
  J. 2007, \pasp, 119, 1186

\bibitem[{{Schneeberger} et~al.(1979){Schneeberger}, {Worden}, \&
  {Beckers}}]{skyb2}
{Schneeberger}, T.~J., {Worden}, S.~P., \& {Beckers}, J.~M. 1979, \pasp, 91,
  530

\bibitem[{Yao et~al.(2012)Yao, Liu, Zhang et~al.}]{1674-4527-12-7-005}
Yao, S., Liu, C., Zhang, H.-T., et~al. 2012, \raa, 12, 772

\end{thebibliography}

\end{document}